**Title:** Investigation of *in vitro* neuronal activity processing using a CMOS-integrated $ZrO_2$-based memristive crossbar

**Authors:** M.N. Koryazhkina[1,2,3], A.V. Lebedeva[3,4], D.D. Pakhomova[3], I.N. Antonov[2,5], V.V. Razin[3], E.D. Budylina[3], A.I. Belov[1,2,5], A.N. Mikhaylov[1,2], A.A. Konakov[6]

**Affiliations:**

[1] Laboratory of Memristor Nanoelectronics, Lobachevsky State University of Nizhny Novgorod, 23/3 Gagarin Ave., 603022 Nizhny Novgorod, Russia

[2] Laboratory of Intellectual Neuromorphic Systems, Lobachevsky State University of Nizhny Novgorod, 23/3 Gagarin Ave., 603022 Nizhny Novgorod, Russia

[3] Research Institute of Neuroscience, Lobachevsky State University of Nizhny Novgorod, 23/7 Gagarin Ave., 603022 Nizhny Novgorod, Russia

[4] Privolzhsky Research Medical University, 10/1 Minin and Pozharsky Sq., 603950 Nizhny Novgorod, Russia

[5] Educational Electronics Design Center of Lobachevsky University, Lobachevsky State University of Nizhny Novgorod, 25/1 Gagarin Ave., 603057 Nizhny Novgorod, Russia

[6] Laboratory of Materials for Quantum Technologies, Lobachevsky State University of Nizhny Novgorod, 23/3 Gagarin Ave., 603022 Nizhny Novgorod, Russia

**Abstract:** The influence of the epileptiform neuronal activity on the response of a CMOS-integrated $ZrO_2$-based memristive crossbar and its conductivity was studied. Epileptiform neuronal activity was obtained *in vitro* in the hippocampus slices of laboratory mice using 4-aminopyridine experimental model. Synaptic plasticity of the memristive crossbar induced by epileptiform neuronal activity pulses was detected. Qualitatively, the results obtained in the case of normal (without pathologies) and epileptiform neuronal activity with and without noise coincide. For quantitative analysis, the value of the relative change in synaptic weight has been calculated for such important biological mechanisms of synapses as paired-pulse facilitation/depression, post-tetanic potentiation/depression, and long-term potentiation/depression. It has been shown that average value of the relative change in synaptic weight and it's are smaller mainly in the case of epileptiform neuronal activity pulses. An effect of the influence of noise included in the neuronal activity was found, which consists in the fact that the current response of the memristive crossbar is smaller in the presence of noise. The results of this study can be used in the development of new generation hardware-implemented computing devices with high performance and energy efficiency for the tasks of restorative medicine and robotics. In particular, using these results, neurohybrid devices can be developed for processing epileptiform activity in real time and for its suppression.

**1. Introduction**

Epilepsy is a severe chronic disease characterized by spontaneous and uncontrollable seizures. Epilepsy significantly affects the quality and expectancy of patient's life [1]. Currently, the most common treatment tactics are drug therapy and surgery, but about 30% of patients do not respond to treatment with anticonvulsants, falling into the drug-resistant epilepsy category patients. Surgical resection is one of the most reliable methods to the stop occurrence of seizures, but its capabilities are limited. It is not possible to remove the entire zone of seizure occurrence frequently and surgical intervention can lead to additional complications and the appearance of new epileptogenic zones due to damage to brain structures.

A promising treatment for epilepsy is neurostimulation. It involves the use of electrical or magnetic currents in the brain to change and stop the pathological neuronal activity [2]. By acting directly on a specific area, neurostimulation can regulate the symptoms manifestation, stop attacks or change the frequency and duration. This procedure remains reversible and minimally invasive. In addition, neurostimulation allows avoiding many of the serious side effects associated with medications or surgery. According to modern concepts, epilepsy is a network disease involving many brain structures whose activities are interconnected. This makes neurostimulation a possible effective method for interrupting pathological hypersynchronous activity in the interconnected neuronal network of the brain [3]. However, existing stimulation methods are generally not adaptive and flexible enough [4–9]. Devices used in clinical practice apply stimulating signals of constant frequency and amplitude without feedback. This significantly increases negative effects and reduces the effectiveness of stimulation [10].

Thus, the development of new methods and wearable devices, namely neurointerfaces or brain-computer interfaces, for stopping and suppressing epileptiform neuronal activity are urgent tasks in medicine. Recently, research related to the development of neurointerfaces has focused on the use of memristor or memristive device [11–16]. Memristor are considered as promising elements in such devices. They are capable of gradually changing their conductivity and, as a result, demonstrate a number of functions inherent in biological neurons [17–21], such as short-term or long-term potentiation/depression, corresponding to an increase/decrease in the efficiency of information transfer between two nerve cells as a result of external influence [22]. Excitable neurons in the brain respond differently to stimuli, just as applying voltage pulses of certain amplitude, duration, and frequency can cause a change in the conductivity of the memristor or its resistive state, which is reflected in a change in the value of the current flowing through it. The properties of memristors can help in the analysis of epileptiform neuronal activity and become a component of new neurointerfaces for processing epileptiform signals and their further use in wearable devices for stopping seizures [23, 24]. Moreover, the coauthors of this paper have shown that memristors can be used in a hybrid neuromorphic system that makes it possible to stimulate areas of the hippocampus bypassing damaged areas [25, 26], as well as memristor-based neural network can be used to predict epileptiform activity before its actual occurrence [27]. Nevertheless, first of all, it is necessary to determine the influence of the parameters of epileptiform neuronal activity on the resistive state of the memristor. It should also be noted that adequate studies should be performed using memristors that demonstrate significant endurance to repeated changes in the resistive state.

In this paper, we investigate the processing of neuronal activity by a memristive crossbar based on a Ta/ZrO$_2$(Y)/Pt stack, assessing its functional properties and changing the resistive state of the device in response to epileptiform neuronal activity obtained *in vitro* in the hippocampus slices of laboratory mice using 4-aminopyridine experimental model. Moreover, the work investigates the influence of noise included in the neuronal activity on the response of the memristive crossbar. The point is that the effect of noise on a memristor leads to the emergence of noise-induced phenomena, where noise plays a constructive or positive role (see, for example, works devoted to the detection of such phenomena in memristors as stochastic resonance [28, 29], noise enhanced stability [30, 31], Random Telegraph Noise observations [32, 33], etc.).

**2. Materials and methods**

## 2.1. Ethics statement

In this study, the protocols of all experiments were reviewed and confirmed by the Bioethics Committee of the Institute of Biology and Biomedicine National Research of the Lobachevsky State University of Nizhny Novgorod.

## 2.2. Electrophysiological experiments

The objects of biological experiments were the surviving hippocampus slices of c57bl/6 mice aged from 3 to 7 months, obtained at the UNN Center for Genetic Collections of Laboratory Animals. All animals received food and water *ad libitum* and were kept on a 12-hour light cycle.

Local field potentials (LFP) were recorded on 380 μm thick hippocampus slices obtained using a vibratome Microm HM 650 B (Thermo Fisher Scientific). Hippocampal mice slice preparation was carried out in accordance with the method described in [25]. After preparation, the slices were placed in the chamber of an Olympus BX51W1 electrophysiological microscope to record neuronal activity (LFP).

A schematic representation of the experimental setup for recording neuronal activity is shown in Figure 1. The recording was carried out in the CA1 *str.radiatum* field of the hippocampus using a glass microelectrode with a resistance of 3-7 MΩ at electrical stimulation of the *dentate gyrus* (DG) with a stimulus duration of 500 μs and different stimulus amplitudes (100, 200, 300, 400, 500, 1000 μA). The stimulation was implemented using an industrial DS3 Isolated Current Stimulator (Digitimer Ltd). Recording of neuronal activity was performed using biological signal amplifier Axon 700B and multifunction data acquisition device (DAQ) National Instruments USB-6211.

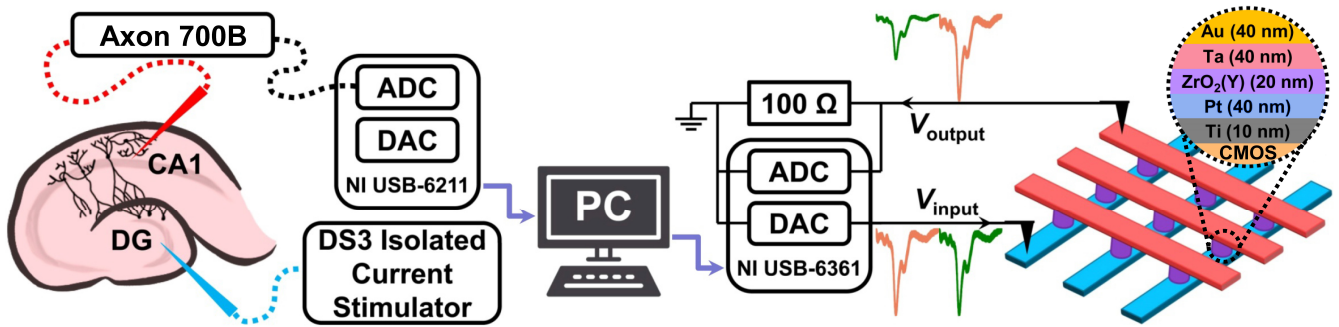

**Figure 1.** Schematic representation of the experimental setup for recording neuronal activity (left) and investigating the response of a memristive crossbar based on a Ta/ZrO$_2$(Y)/Pt stack to this neuronal activity (right): the *dentate gyrus* (DG) is stimulated using the industrial DS3 Isolated Current Stimulator (Digitimer Ltd); the response to stimulation was recorded in the *str.radiatum* field of the CA1 region using the biological signal amplifier Axon 700B and DAQ National Instruments USB-6211; then the response to electrical stimulation, namely the LFP, was sent to a personal computer and underwent pre-processing; the pre-processed neuronal activity was used to prepare the input control signal (see Figure 2 for explanation); the resulting input control signal was applied to a CMOS-integrated memristive crossbar using the Everbeing EB-6 probe station and the multifunction DAQ National Instruments USB-6361; the response of the memristive crossbar to input control signal was recorded as a voltage drop on a load resistor and was sent to a personal computer

Fifty recordings of LFP containing epileptiform neuronal activity were obtained, which was experimental modeled using the chemoconvulsant 4-aminopyridine at a concentration of 1 μM. The obtained LFP recordings were pre-processed as follows. First, the artifact corresponding to the stimulus was removed, then the curves were normalized to the baseline, after that the curves were filtered using a Gaussian filter, and finally the curves were averaged. The curves were then scaled so that their amplitudes were close in order of magnitude to the resistive switching voltages. It turned out that after pre-processing, the curves obtained by stimulation with different amplitudes coincide. It should be noted that for the experiment with noisy neuronal activity, the pre-processing steps corresponding to filtering and averaging of curves were skipped. To determine the noise parameters, the filtered and averaged curve was subtracted from the noisy neuronal activity. It turned out that this noise has a Gaussian distribution with zero mean and a standard deviation of 0.037 V for curves normalized to unity, i.e. for which the amplitude is 1 V.

### 2.3. Memristive crossbar preparation

The studies were performed using a memristive device in a 32 × 8 crossbar-array integrated within the framework of the BEOL (back-end-of-line) process with 0.35 μm complementary metal-oxide-semiconductor (CMOS) devices (see explanation about integration of memristors with CMOS in [34]). The memristive crossbar consisted of the following layers obtained by magnetron sputtering: Au (40nm)/Ta (40nm)/$ZrO_2$(12 % mol. $Y_2O_3$) (20nm)/Pt (40nm)/Ti (10nm). Detailed information about the fabrication technology of the investigated device can be found in [31]. The active area of memristive crossbar was 10 × 10 μm$^2$.

### 2.4. Methods of memristive crossbar investigation

The research was carried out using the Everbeing EB-6 probe station and the multifunction DAQ National Instruments USB-6361 (see Figure 1). In order for current to flow through the memristive crossbar, the CMOS device was opened by applying a voltage of +3.3 V. In all experiments the sign of voltage across memristive crossbar corresponded to the potential of Pt electrode relative to the potential of Ta electrode. With such a measuring circuit, switching from a low-resistance state (LRS) to a high-resistance state (HRS), so-called RESET-process, was achieved by applying a voltage of positive polarity, and vice versa, from a HRS to a LRS, so-called SET-process, by applying a voltage of negative polarity. And, accordingly, depression was achieved by applying a series of voltage pulses of positive polarity, and potentiation was achieved by applying a series of voltage pulses of negative polarity.

The response of the memristive crossbar to input control signal, namely epileptiform neuronal activity or rectangular pulse sequences for testing resistive switching endurance, was recorded as a voltage drop on a load resistor. All studies were carried out on the same memristive crossbar.

The pre-processed neuronal activity was used to prepare the input control signal. This signal consisting of a one high/low resistance switching pulse of positive/negative polarity and a series of 100 pulses, each representing neuronal activity of negative/positive polarity with a fixed amplitude $A$. The resistive state of the memristive crossbar was read after each pulse. A schematic representation of one cycle of the input control signal is shown in Figure 2. The parameters of switching pulses and neuronal activity pulses (NAP) are shown in Figure 2. The value of $A$ in the potentiation experiments was -0.7 – -0.9V, and in the depression experiments $A$ = +0.7 – +1.1 V. For each case of the $A$, the memristive crossbar response to 100 cycles of input control signal application was recorded.

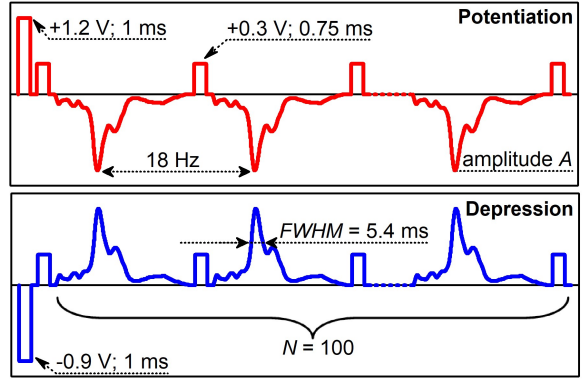

**Figure 2.** Schematic representation of one cycle of the input control signal with epileptiform neuronal activity

The response of the memristive crossbar to the input control signal was processed in the same way as was done earlier in the experiments to determine the influence of normal, i.e. without pathologies, *in vitro* neuronal activity of the hippocampus slices of laboratory mice on the response of the same memristive crossbar [35].

Additionally, for the purpose of quantitative analysis of the obtained results, the value of the relative change in synaptic weight was calculated for different NAP. The value of the relative change in synaptic weight was determined as $K_n = ((|I_n - I_1|) / I_1) \cdot 100\%$ [36], where $I_n$ is the value of the current flowing through the memristive crossbar at the reading after applying the $n^{th}$ NAP, $I_1$ is the value of the current flowing through the memristive crossbar at the reading after applying the first NAP.

### 3. Experimental results and discussion

### 3.1. Preliminary characterization of memristive crossbar

Figure 3 shows the 50 $I - V$ curves of the memristive crossbar. The curves demonstrate a gradual change in the resistive state and sufficient stability of the resistive switching parameters, namely, currents in resistive states and resistive switching voltage. Note that the selected range of the $A$ covers the range of voltages at which changes in the resistive state are observed on the $I - V$ curves.

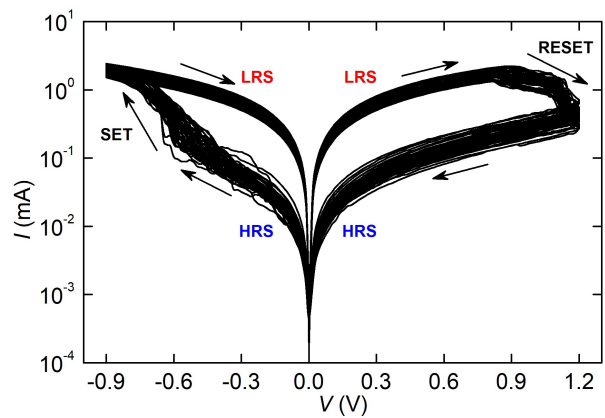

**Figure 3.** The 50 $I - V$ curves of memristive crossbar based on a Ta/ZrO$_2$(Y)/Pt stack

Figure 4 shows the results of the study of the memristive crossbar endurance to multiple resistive switching. It is shown that the device does not exhibit degradation up to 10 000 resistive switching cycles. It should be noted that the currents in resistive states are highly stable from cycle to cycle (C2C) – the coefficient of variation, in some sense equivalent to scatter, is 1.1% for currents in the

LRS and 11% for currents in the HRS. The ratio of the minimum value of current in the LRS to the maximum value of current in the HRS is 5, and on average, the ratio of currents in the LRS and HRS is 8.

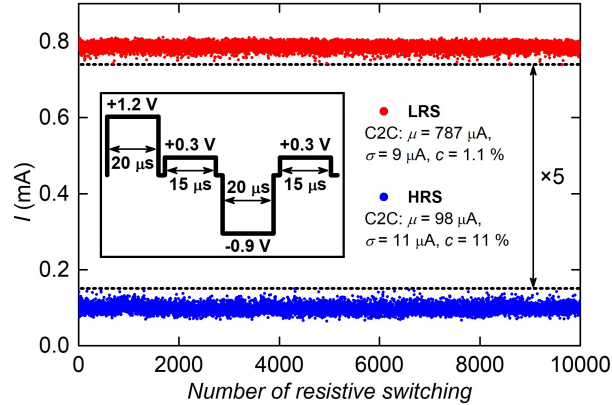

**Figure 4.** Endurance of memristive crossbar based on a Ta/ZrO$_2$(Y)/Pt stack. The figure shows the average value of the current ($\mu$), its standard deviation ($\sigma$) and the coefficient of variation ($c = \sigma / \mu$). Inset: schematic representation of an endurance test

### 3.2. Response of a memristive crossbar to epileptiform neuronal activity

Figures 5 and 6 shows the dependencies of the current obtained during reading on the number of readings. Figure 5 shows the results for the first five cycles of applying an input control signal and Figure 6 shows the curves averaged over 100 cycles of applying an input control signal. For almost all $A$, a gradual change in the resistive state is observed when applying a series of NAP's both with and without noise.

First of all, it can be noted that at $|A| = 0.7 – 0.9$ V (except for $A = -0.9$ V, which will be discussed below), the current values in the case of NAP with noise are less than NAP without noise. The explanation of the effect may be related to the constructive role of noise, namely, the noise enhanced stability phenomenon. However, with the $|A|$ used in this study, the standard deviation of noise varies from 0.026 to 0.041 V, which in terms of noise intensity corresponds to $3.4 \cdot 10^{-8}$ and $8.4 \cdot 10^{-8}$ V$^2 \cdot$s. The obtained values are significantly greater than those considered in the study [31]. Therefore, a more detailed explanation of the effect discovered in this paper requires comprehensive studies of the influence of a widely varying intensity of the noise, which is part of neuronal activity, on the response of the memristive crossbar. At $A = +1.0$ and $+1.1$ V, the current values in the case of a NAP with and without noise are practically the same, which indicates the predominant influence of the amplitude value.

The case of $A = -0.9$ V should be considered separately. It is evident that with each subsequent NAP without noise, a decrease in conductivity is observed, despite the fact that at smaller $|A|$ the opposite behavior is observed. Studies performed with normal (without pathologies) neuronal activity demonstrated a decrease in conductivity only at $A < -1.0$ V (not given in the study). It is assumed that at such values of $A$, the effect of heating the filament, aimed at its rupture, is more significant than the effect of the electric field, aimed at restoration of the filament. In this regard, studies at $A$ lower than -0.9 V were not performed.

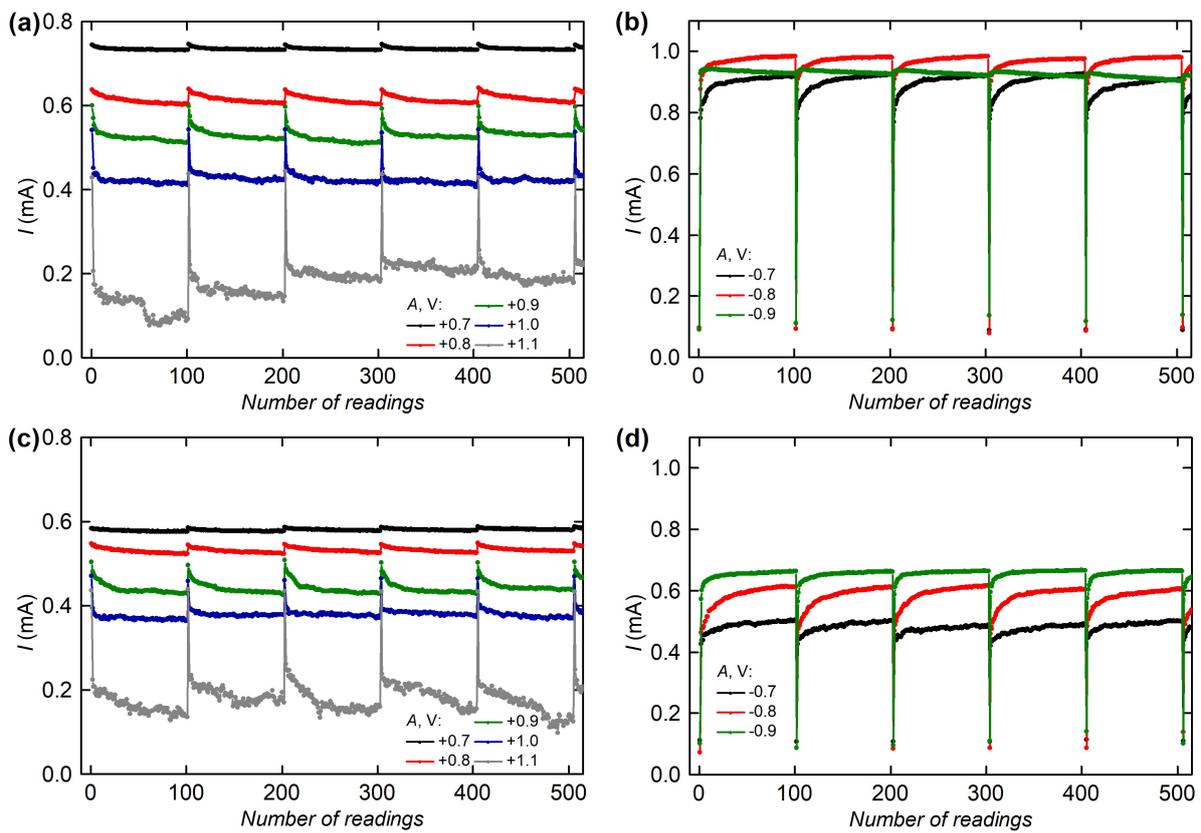

**Figure 5.** Dependencies of the current obtained during reading on the number of readings for the first five cycles of applying an input control signal with $A > 0$ **(a, c)** and $A < 0$ **(b, d)**. Data obtained using epileptiform NAP **(a, b)** and noisy epileptiform NAP **(c, d)**

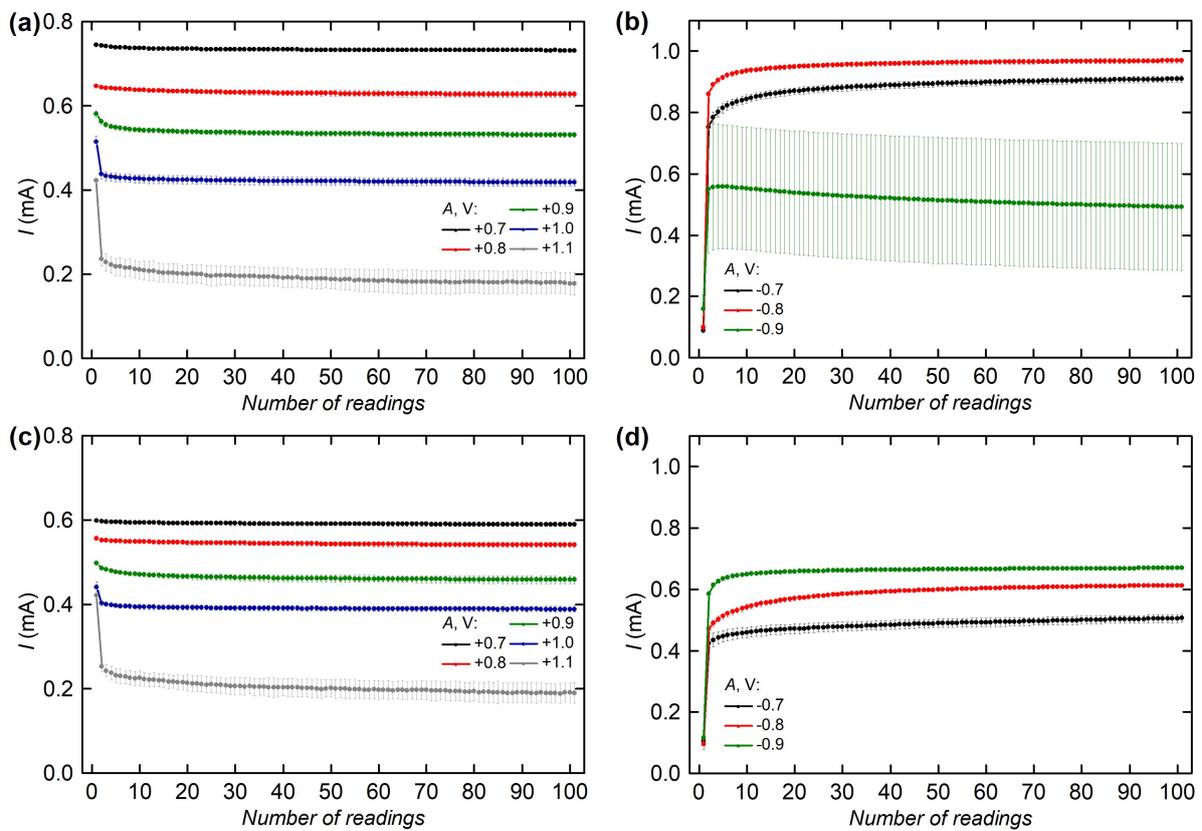

**Figure 6.** Dependencies of the averaged current obtained during reading on the number of readings. Data obtained at input control signal with $A > 0$ **(a, c)** and $A < 0$ **(b, d)** using epileptiform NAP **(a, b)** and noisy epileptiform NAP **(c, d)**

Figures 7 and 8 shows the averaged current response of the memristive crossbar to a specific pulse in a series of NAP.

At $A$ = +0.7 and +0.8 V the curves were almost indistinguishable from each other. Increasing $A$ value to +0.9 V resulted in the response amplitude decreasing with each new pulse from the NAP series. At $A$ = +1.0 V, differences in shape between the current response of the memristive crossbar to the 1$^{st}$ NAP and subsequent NAP's begin to be observed. With a further increase in $A$ to +1.1 V, the differences in shape become more obvious, and the difference in the amplitude of the current response to the NAP in the series becomes more significant. It should be noted that with the increase of $A$ the maximum of the curves decreases.

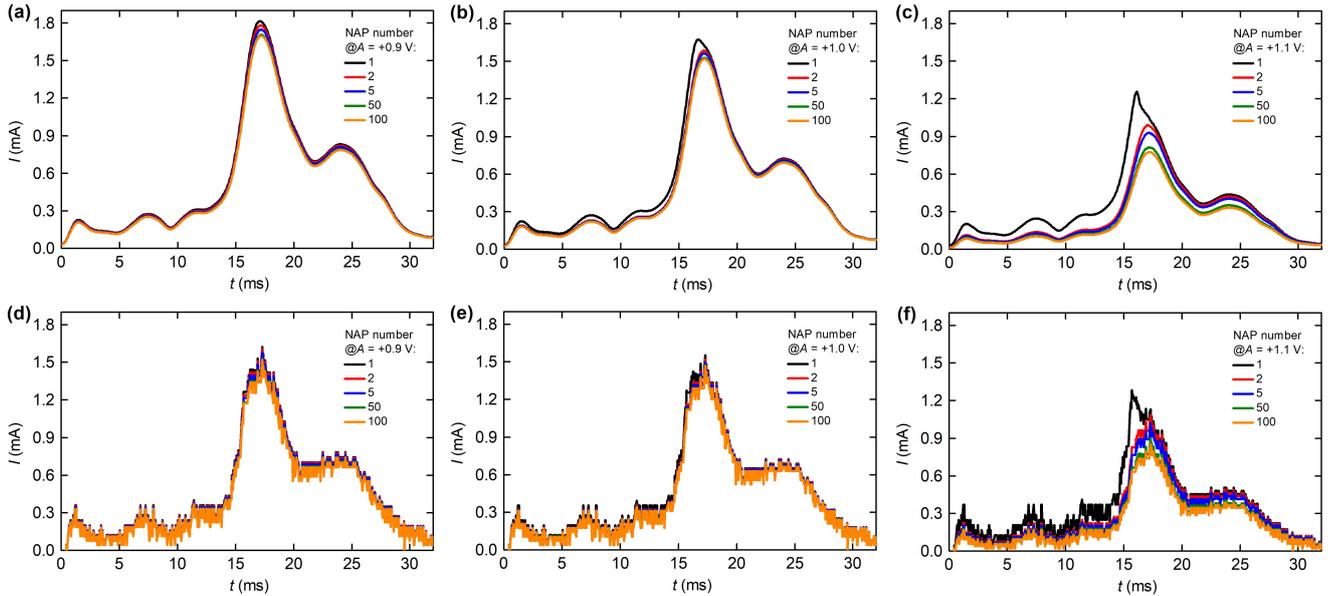

**Figure 7.** Averaged current responses of the memristive crossbar to a specific pulse in a series of NAP with different $A$, V: +0.9 **(a, d)**, +1.0 **(b, e)**, and +1.1 **(c, f)**. Data obtained using epileptiform NAP **(a, b, c)** and noisy epileptiform NAP **(d, e, f)**

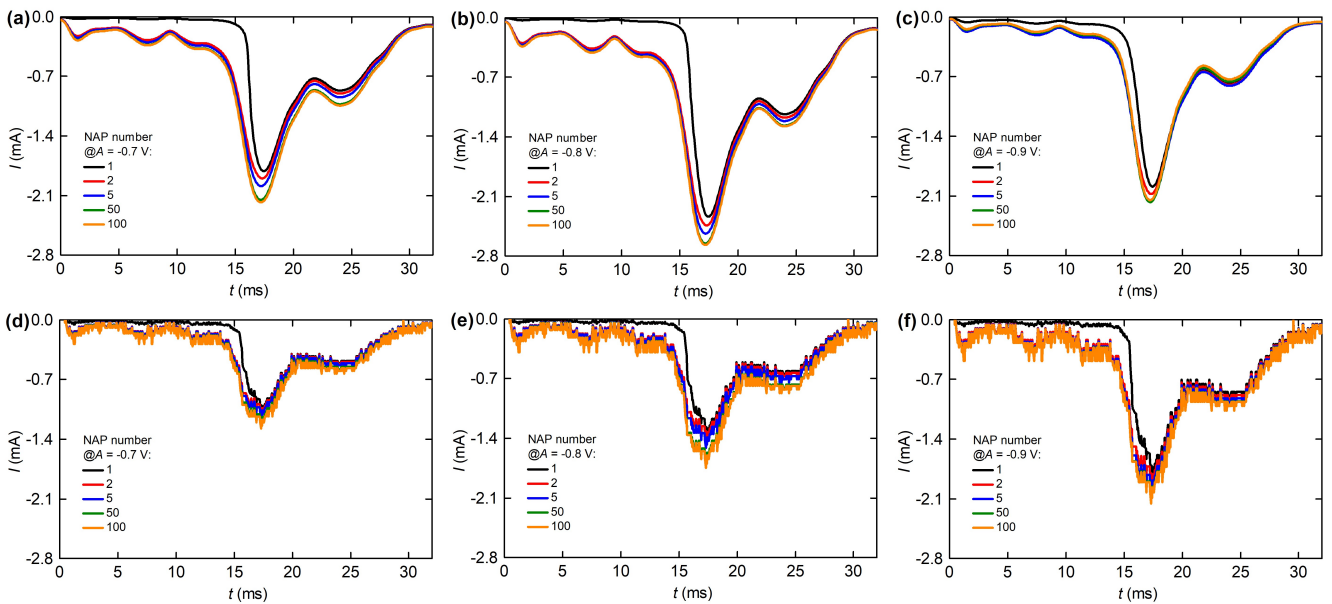

**Figure 8.** Averaged current responses of the memristive crossbar to a specific pulse in a series of NAP with different $A$, V: -0.7 **(a, d)**, -0.8 **(b, e)**, and -0.9 **(c, f)**. Data obtained using epileptiform NAP **(a, b, c)** and noisy epileptiform NAP **(d, e, f)**

In the case of $A < 0$, differences were observed immediately at -0.7 V. Further decrease in $A$ to -0.8 V led to a less noticeable effect. As already noted above, in the case of -0.9 V without noise, a decrease in the amplitude of the current response was observed with each subsequent NAP.

The results obtained in all three experiments (epileptiform NAP with and without noise, as well as normal NAP [35]) are in qualitative agreement. The results of the quantitative analysis are discussed in more detail below.

Quantitative analysis of the obtained results was performed by determining the value of the relative change in synaptic weight for different NAP (Figure 9).

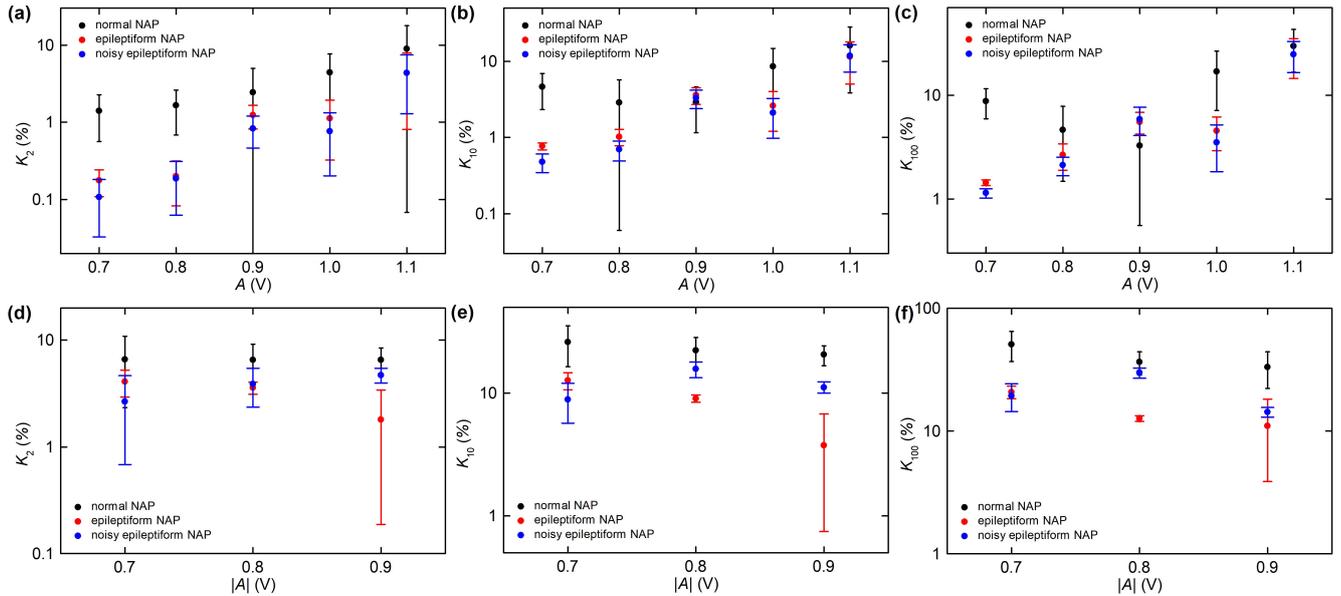

**Figure 9.** Dependencies of the value of the relative change in synaptic weight $K_2$ **(a, d)**, $K_{10}$ **(b, e)**, and $K_{100}$ **(c, f)** on a value of $A$. Data obtained using normal [35], epileptiform, and noisy epileptiform NAP with $A > 0$ **(a, b, c)** and $A < 0$ **(d, e, f)**

First, let us consider the case of $A > 0$. With increasing $A$, the values of $K_2$, $K_{10}$, and $K_{100}$ tend to increase. On average, these values are larger mainly in the case of normal NAP. At the same time, the presence of noise leads to a decrease in the average values of $K_2$, $K_{10}$, and $K_{100}$. The scatter of these values in the case of epileptiform NAP is mainly smaller than in the case of normal NAP. It should be emphasized that on average all three values in the case of epileptiform NAP both with noise and without noise increase up to $A < +1.0$ V, then at $A = +1.0$ V a tendency to decrease is observed, and after that at $A = +1.1$ V a tendency to increase is observed again.

Now, let us consider the case of $A < 0$. With increasing $|A|$, on average, the values of $K_2$, $K_{10}$, and $K_{100}$ tend to decrease for the cases of normal and epileptiform NAP without noise. For the case of epileptiform NAP with noise, the value of $K_2$ tends to increase, while the values of $K_{10}$ and $K_{100}$ show a clearly defined maximum. On average, the $K_2$, $K_{10}$, and $K_{100}$ are larger mainly in the case of normal NAP. The scatter of these values in the case of epileptiform NAP is mainly smaller than in the case of normal NAP.

It should be noted that the $K_2$ value corresponded to the change in synaptic weight during paired-pulse facilitation/depression, $K_{10}$ – during post-tetanic potentiation/depression, $K_{100}$ – during long-term potentiation/depression. These phenomena are referred to as short-term and long-term synaptic plasticity and represent a set of important behavioral mechanisms of biological synapses, which are believed to underlie learning and memory.

## 4. Conclusions

The influence of the epileptiform neuronal activity on the response of a CMOS-integrated memristive crossbar based on a Ta/ZrO$_2$(Y)/Pt stack was studied. Preliminary characterization of the memristive crossbar showed that the device does not exhibit degradation up to 10 000 resistive switching cycles, and the currents in resistive states are highly stable from one resistive switching cycle to another cycle.

Epileptiform neuronal activity was obtained *in vitro* in the hippocampus slices of laboratory mice using 4-aminopyridine experimental model.

Synaptic plasticity of the memristive crossbar induced by epileptiform neuronal activity pulses was detected. It is shown that the amplitude of neuronal activity can be used as a parameter for controlling the synaptic plasticity properties of a memristive crossbar.

In this study, a qualitative and quantitative comparison of the responses of a CMOS-integrated ZrO$_2$(Y)-based memristive crossbar to normal (without pathologies) and epileptiform (with and without noise) neuronal activity was carried out. Qualitatively, the results obtained in the case of different neuronal activity coincide.

For quantitative analysis, the value of the relative change in synaptic weight has been calculated for such important behavioral mechanisms of biological synapses as paired-pulse facilitation/depression, post-tetanic potentiation/depression, and long-term potentiation/depression. It has been shown that average value of the relative change in synaptic weight and it's are smaller mainly in the case of epileptiform neuronal activity pulses.

An effect of the influence of noise included in the neuronal activity was found, which consists in the fact that the current response of the memristive crossbar is smaller in the presence of noise. However, this result requires more thorough and comprehensive studies of the influence of noise intensity superimposed on neuronal activity on the response of the memristive crossbar.

The results of this study can be used in the development of new generation hardware-implemented computing devices with high performance and energy efficiency for the tasks of restorative medicine and robotics. In particular, using these results, neurohybrid devices can be developed for processing epileptiform activity in real time and for its suppression.

### CRediT authorship contribution statement

**M.N. Koryazhkina:** Conceptualization, Investigation, Formal analysis, Data curation, Visualization, Writing – original draft, Funding acquisition. **A.V. Lebedeva:** Conceptualization, Methodology, Formal analysis, Writing – original draft, Project administration. **D.D. Pakhomova:** Investigation, Writing – original draft. **I.N. Antonov:** Resources. **V.V. Razin:** Methodology. **E.D. Budylina:** Writing – original draft. **A.I. Belov:** Methodology. **A.N. Mikhaylov:** Writing - Review & Editing, Funding acquisition. **A.A. Konakov:** Writing - Review & Editing.

### Declaration of competing interest

The authors declare that they have no known competing financial interests or personal relationships that could have appeared to influence the work reported in this paper.

**Data availability statement**

The data that support the findings of this study are available from the corresponding author upon reasonable request.

**Acknowledgments**

The study was supported by the Russian Science Foundation (Grant No. 24-21-00440, https://rscf.ru/project/24-21-00440/).